\documentclass[twocolumn,letterpaper]{IEEEAerospaceCLS}

\usepackage[]{graphicx,multirow}    % We use this package in this document
\usepackage{svg}
\usepackage[fleqn]{amsmath}
\usepackage{bm}
\newcommand{\ignore}[1]{}  % {} empty inside = %% comment
\graphicspath{ {figs/} } % relative path to figs here
\usepackage{acronym}
\usepackage{subfig}
\usepackage{balance}
\usepackage{lipsum} 
\usepackage{booktabs}
\usepackage{cite}
\usepackage{float}

\makeatletter
\def\endthebibliography{%
	\def\@noitemerr{\@latex@warning{Empty `thebibliography' environment}}%
	\endlist
}
\makeatother

\usepackage{adjustbox}
\usepackage{array}

\newcolumntype{R}[2]{%
	>{\adjustbox{angle=#1,lap=\width-(#2)}\bgroup}%
	l%
	<{\egroup}%
}
\begin{document}
	
	%%%%%%%%%%%%%%%%%%%%%%%%%%%%%%%%%%%%%%
	%% Acronyms
	%%%%%%%%%%%%%%%%%%%%%%%%%%%%%%%%%%%%%%
	\acrodef{NASA}{National Aeronautics and Space Administration}
	\acrodef{JPL}{Jet Propulsion Laboratory}
	%%%%%%%%%%%%%%%%%%%%%%%%%%%%%%%%%%%%%%
	%% Title
	%%%%%%%%%%%%%%%%%%%%%%%%%%%%%%%%%%%%%%
	
	\title{How to Deploy a 10-km Interferometric Radio Telescope on the Moon with Just Four Tethered Robots}
	%%%%%%%%%%%%%%%%%%%%%%%%%%%%%%%%%%%%%%
	%% Authors
	%%%%%%%%%%%%%%%%%%%%%%%%%%%%%%%%%%%%%%

	\author{%
		Patrick McGarey, Issa A. Nesnas, Adarsh Rajguru, Matthew Bezkrovny,\\ Vahraz Jamnejad, Jim Lux, Eric Sunada, Lawrence Teitelbaum\\ 
		Jet Propulsion Laboratory, California Institute of Technology\\
		\textit{first}.\textit{last}@jpl.nasa.gov
		\and
		Alexander Miller, Steve W. Squyres\\ 
		Blue Origin Enterprises\\
		AMiller2, SSquyres@blueorigin.com
		\and
		Gregg Hallinan\\ 
		California Institute of Technology\\
		gh@astro.caltech.edu
		\and
		Alex Hegedus\\ 
		University of Michigan\\
		alexhege@umich.edu
		\and
		Jack O. Burns\\ 
		University of Colorado Boulder\\
		jack.burns@colorado.edu
		% Copyright.
		\thanks{\footnotesize{\textcopyright} 2021. All rights reserved.}     
	}

	%% FOR ARXIV
 	%This is the author’s version of the work. It is posted here for your personal use. Not for redistribution. The definitive Version of Record was published in

	%%%%%%%%%%%%%%%%%%%%%%%%%%%%%%%%%%%%%%
	%% Setup
	%%%%%%%%%%%%%%%%%%%%%%%%%%%%%%%%%%%%%%
	
	\maketitle
	\thispagestyle{plain}
	\pagestyle{plain}
	
	%%%%%%%%%%%%%%%%%%%%%%%%%%%%%%%%%%%%%%
	%% Abstract
	%%%%%%%%%%%%%%%%%%%%%%%%%%%%%%%%%%%%%%
	
	\begin{abstract}
		The Far-side Array for Radio Science Investigations of the Dark ages and Exoplanets (FARSIDE) is a proposed mission concept to the lunar far side that seeks to deploy and operate an array of 128 dual-polarization, dipole antennas over a region of 100 square kilometers. The resulting interferometric radio telescope would provide unprecedented radio images of distant star systems, allowing for the investigation of faint radio signatures of coronal mass ejections and energetic particle events and could also lead to the detection of magnetospheres around exoplanets within their parent star’s habitable zone. Simultaneously, FARSIDE would also measure the `Dark Ages' of the early Universe at a global 21-cm signal across a range of red shifts (z $\sim$50--100). Each discrete antenna node in the array is connected to a central hub (located at the lander) via a communication and power tether. Nodes are driven by cold-operable electronics that continuously monitor (for 5 or more years) an extremely wide-band of frequencies (200 kHz to 40 MHz), which surpass the capabilities of Earth-based telescopes by two orders of magnitude. Achieving this ground-breaking capability requires a robust deployment strategy on the lunar surface, which is feasible with existing, high-TRL technologies (demonstrated or under active development) and is capable of delivery to the surface on next-generation commercial landers, such as Blue Origin’s Blue Moon Lander. This paper presents an antenna packaging, placement, and surface deployment trade study that leverages recent advances in tethered mobile robots under development at NASA’s Jet Propulsion Laboratory, which are used to deploy a flat, antenna-embedded, tape tether with optical communication and power transmission capabilities. We investigate the feasibility of deploying 4 separate 12-km tethers, each with 64 remotely powered electronics nodes, using just four rovers that trace out a large spiral pattern that is over 10-km in diameter, where each arm in the spiral is precisely laid out and aligned with respect to a global heading (NSEW) to provide dual polarization across the entire array. Further, we provide a detailed design for an instrument/antenna-embedded tether, detail its accommodation within and deployment from a two-wheeled rover, and show how the entire system can be packaged and deployed from a commercial lunar lander system.
	\end{abstract}
	
	%%%%%%%%%%%%%%%%%%%%%%%%%%%%%%%%%%%%%%
	
	\tableofcontents
	
	%%%%%%%%%%%%%%%%%%%%%%%%%%%%%%%%%%%%%%
	
	\begin{figure}[h]
		\centering
		\includegraphics[width=\columnwidth]{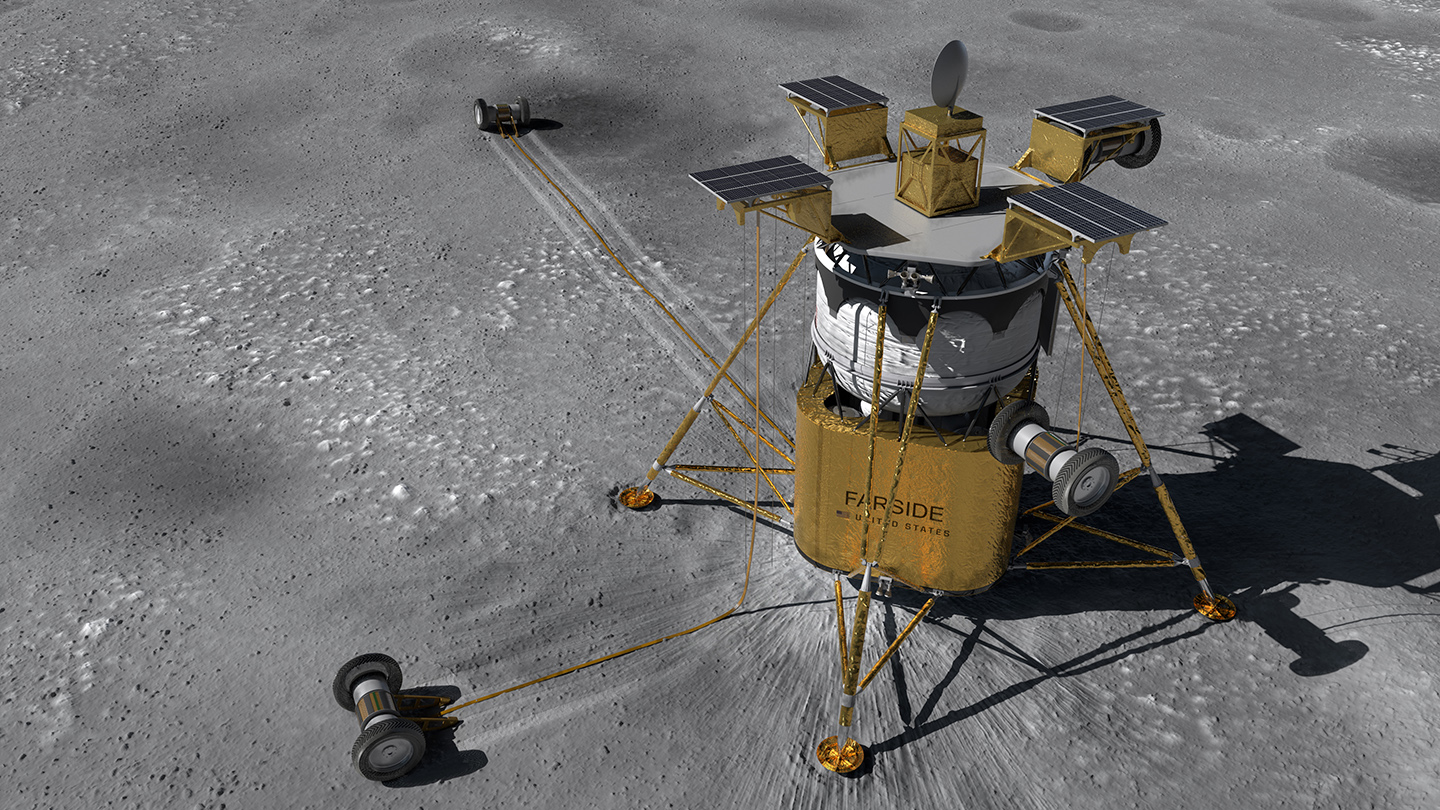}
		\includegraphics[width=\columnwidth]{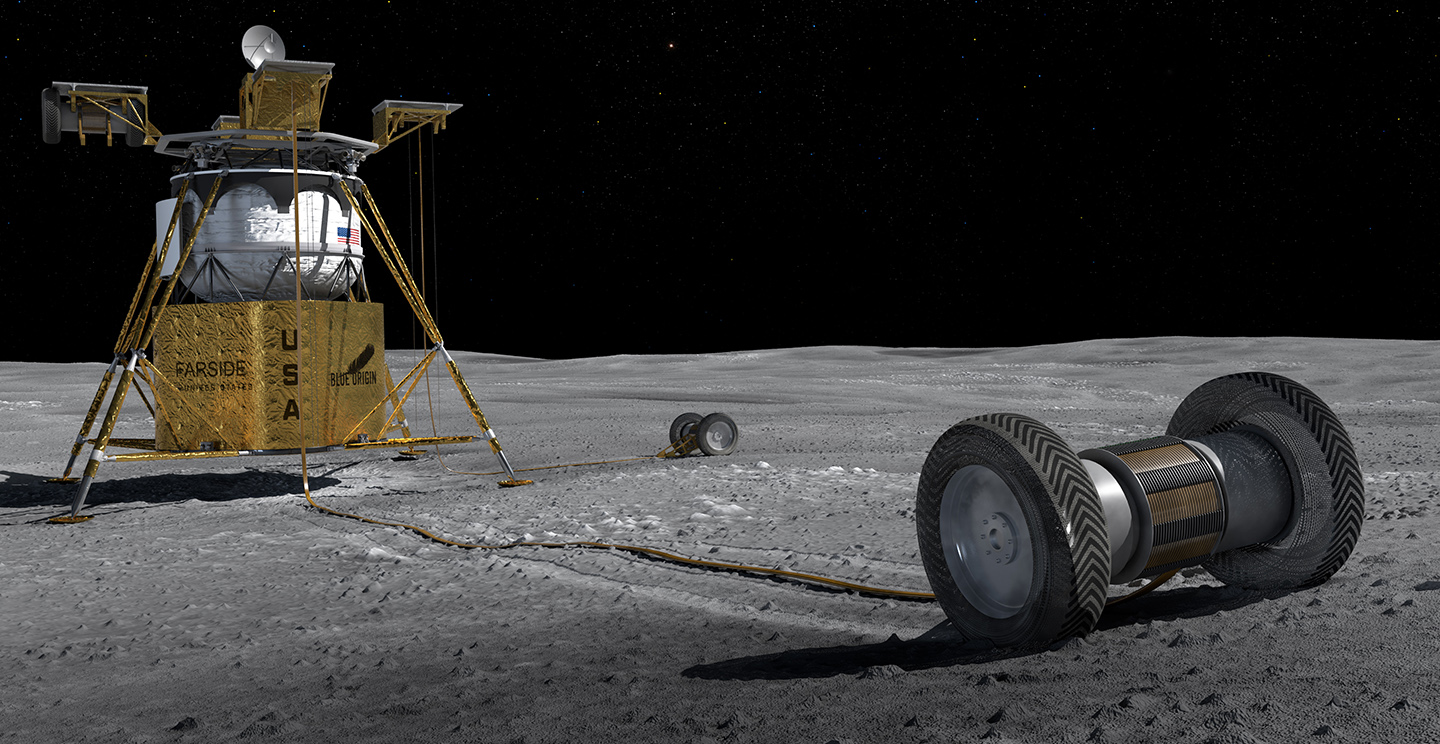}		\caption{\emph{FARSIDE Concept}: Renderings show the initial stages of  tethered-rover egress from a parent lander and the subsequent roll out of an antenna array onto the lunar surface. Antennas and driving electronics are integrated into four 12-km tether rolls, which provide  power and communication to both antenna nodes and deployment rovers. Image credits to XP4D, NASA JPL, and Blue Origin}
		\label{fig:farside-concept}
	\end{figure}

	\section{Introduction}
	\label{sec:Introduction}
	
	The lunar far side is an attractive target for next-generation radio telescopes due to its lack of atmosphere and low light/noise conditions (especially at night). The Far-side Array for Radio Science Investigations of the Dark ages and Exoplanets (FARSIDE) is an interferometer mission concept developed by the University of Colorado Boulder in collaboration with NASA's Jet Propulsion Laboratory (JPL) and California Institute of Technology. FARSIDE is composed of 128 dual-polarization, 100-m, dipole-antenna nodes (256 single-polarization nodes), which are distributed over a 10\,km\,x\,10\,km area in order to detect low-frequency radio emissions from distant star systems, e.g., coronal mass ejections and energetic particle events, as well as the observation of exoplanet magnetospheres. In order to distribute nodes over a large area, while maintaining interconnection and communication to a central lander, we have developed a practical, low-cost deployment strategy using just four tethered rovers that unroll multiple kilometers of antenna-embedded tether in a large spiral pattern. This paper will cover the high-level trades that informed this strategy, some initial scaling parameters, and their impact on the science that can be performed.
	This paper is organized as follows.
	Section\,\ref{sec:background} summarizes the overall FARSIDE mission concept,
	Section\,\ref{sec:trade_study} presents the deployment trade study, 
	Section\,\ref{sec:Conclusion} provides conclusions and future directions for development. 
	
	%%%%%%%%%%%%%%%%%%%%%%%%%%%%%%%%%%%%%%
	
	\vspace{-10px}
	
	\section{Background}
	\label{sec:background}
	
	\begin{figure}[h]
		\centering
		\includegraphics[width=\columnwidth]{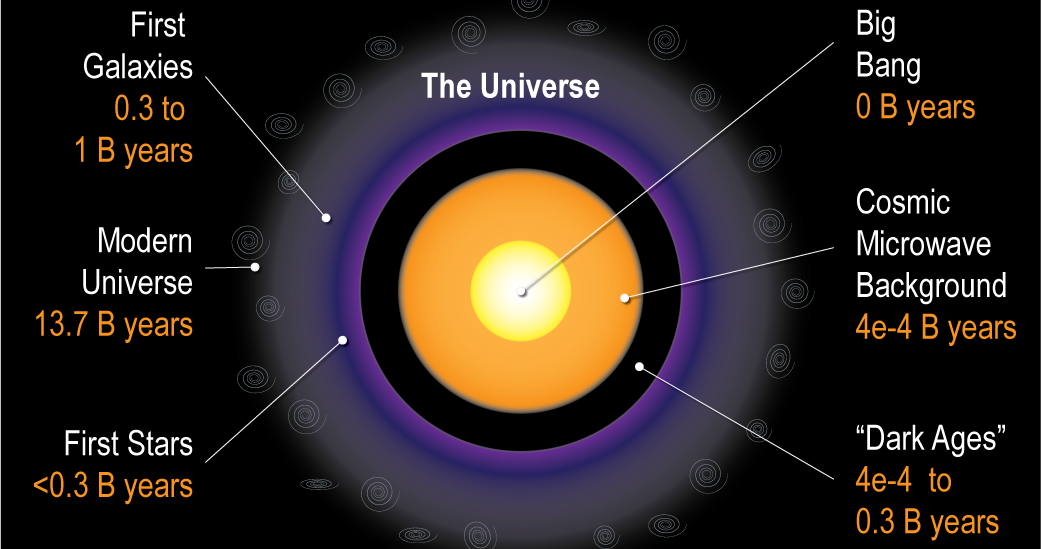}
		\caption{\emph{The History of the Universe}: Regions for which few measurements have been made, including the `Dark Ages' before the formation of stars, galaxies, and later planets are shown. Billions of years are indicated by the letter `B'.}
		\label{fig:universe}
	\end{figure}

	\begin{figure}[h]
		\centering
		\includegraphics[width=\columnwidth]{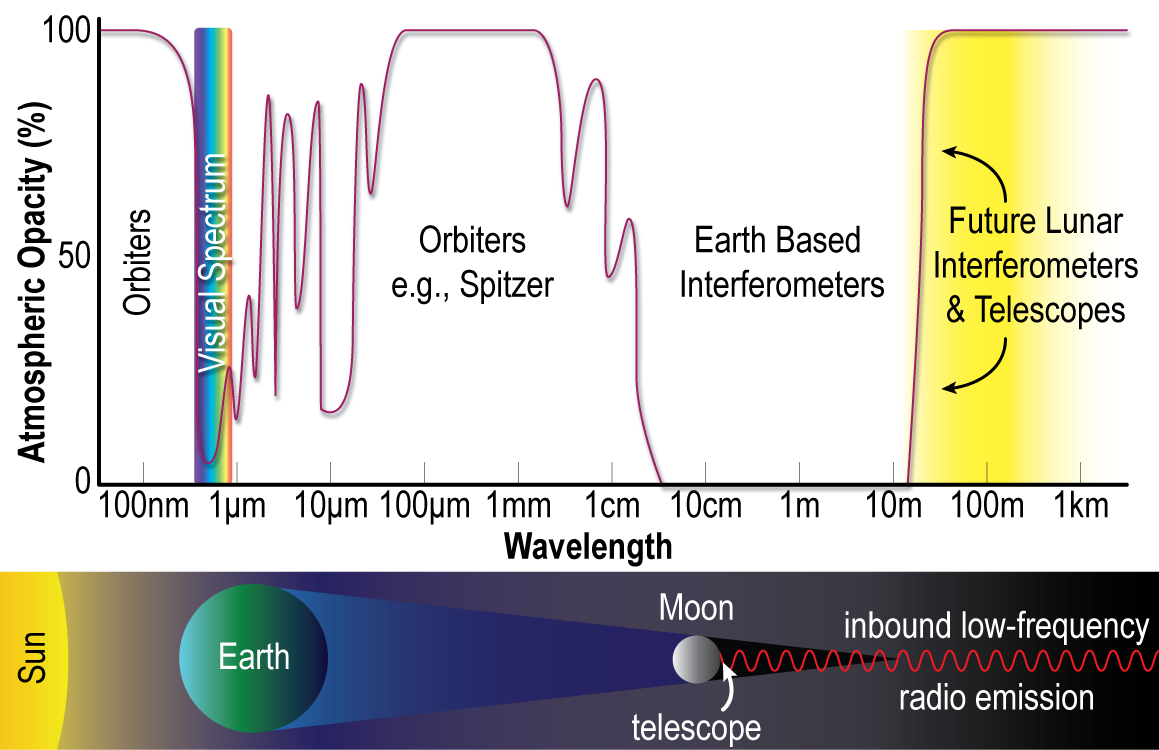}
		\caption{\emph{Science Pull}: (\textit{Top}) the red line shows atmospheric opacity as a function of wavelength to illustrate why certain wavelengths are only visible from outside of Earth. (\textit{Bottom}) Future telescopes and interferometers built on the far side of the Moon will benefit from a drastically noise-reduced environment, from which to make observations of extremely long radio wavelengths coming from very faint sources.}
		\label{fig:spectrum}
	\end{figure}

	On Earth, we are limited to a narrow view of our Universe defined by atmospheric attenuation and the presence of noisy radio emitters. In orbit, measurements can be made over a wider spectrum, with less noise, but such telescopes are still limited by their relative size and distribution to measure very low-frequency radio emissions, such as those emanating from the early ages of our Universe. While orbiting arrays have been proposed, their coordination would be highly complicated and still prone to noise from the Sun and Earth\,\cite{rajan2016space}. Figure\,\ref{fig:universe} illustrates events in the early Universe, including the mostly unobserved `Dark Ages', occurring just after the Big Bang over 13 billion years ago\cite{miralda2003dark}.

\begin{figure}[h]
		\centering
		\includegraphics[width=0.8\columnwidth]{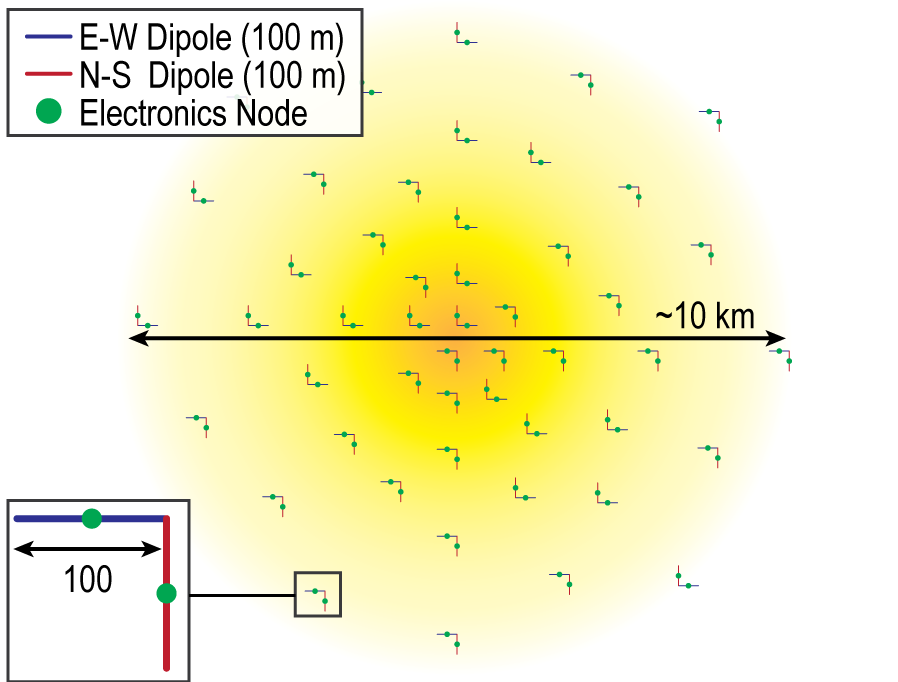}
		\caption{\emph{FARSIDE Node Distribution}: This illustration shows the ideal placement of dual-polarization dipole nodes, which are concentrated in the center and become increasingly more sparse towards the edges.}
		\label{fig:ideal-node-distribution}
	\end{figure}

	In order to fill this knowledge gap, new approaches are needed for next-generation telescopes that can measure very-long wavelengths. Figure\,\ref{fig:spectrum} demonstrates the problem with state-of-the-art implementations, which have, so far, covered  wavelengths shorter than needed to observe the `Dark Age' period prior to the formation of the first stars and galaxies.
	The idea of constructing an interferometric antenna array on the far side of the Moon has been proposed in many iterations for decades\cite{burns1989lunar,burns1990lunar,lazio2009dark,lazio2011radio,burns2011dark}.
	Recently, with the push to return astronauts to the lunar surface in support of future outposts\,\cite{smith2020artemis}, and the development of extreme-terrain planetary rovers\,\cite{nesnas2012axel} that are supported by robust electromechanical tethers\,\cite{mcgarey2020design}, the feasibility of building a radio telescope on the far side of the Moon has been revisited\,\cite{bandyopadhyay2018conceptual,burns2019farside,burns2021lunar}. In particular, the FARSIDE team has worked through the problem of deployment and has formulated a novel approach that can be achieved within the limits of technology in development today. The aim of FARSIDE is to distribute an assortment of dual-polarization, dipole nodes oriented with respect to lunar cardinal directions (NSEW) as shown in Figure\,\ref{fig:ideal-node-distribution}.

	\begin{figure}[h]
		\centering
		\includegraphics[width=\columnwidth]{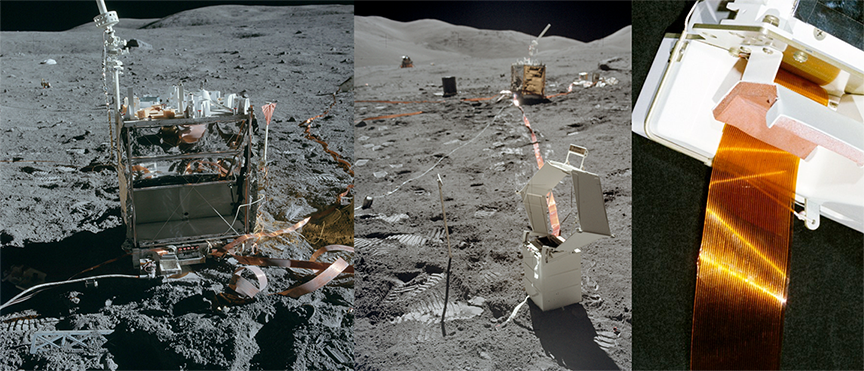}
		\includegraphics[width=\columnwidth]{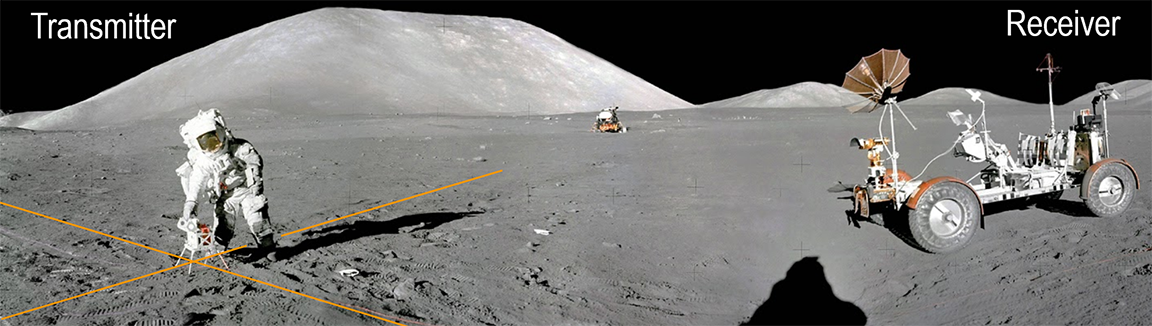}
		\caption{\emph{Apollo 17 Heritage}: (\textit{Top}) the Apollo Lunar Surface Experiments Package (ALSEP) featured long tape tethers deployed on the surface\,\cite{bates1979alsep}. (\textit{Bottom}) The Surface Electrical Properties (SEP) Experiment is the first known surface deployed dipole on another planet or body\,\cite{simmons197315,grimm2018new}.}
		\label{fig:apollo-17-sep}
	\end{figure}

	As shown in Figure\,\ref{fig:apollo-17-sep}, the heritage of deploying antenna arrays outside of Earth goes back to the Apollo era, whose astronauts conducted a number of experiments involving tethers, including a precursor to modern ground penetrating radar (GPR) using a 70-m cross dipole.
	The key takeaway from these experiments, as noted by astronauts, was that, ``Although the cables were fairly stiff, on Earth they tended to lie flat to the ground, held down by their own weight. However, on the Moon, the cables retained loops and bends they had acquired during storage inside the Lunar Module (LM) - loops that stood up from the ground rather like sections of a frozen garden hose.'' Thus, the material selection and storage approaches for FARSIDE tether/antennas are considered as a means to reduce the cable-memory effect and achieve more accurate pattern layouts on the lunar surface.

	%%%%%%%%%%%%%%%%%%%%%%%%%%%%%%%%%%%%%%

	\section{Trade Study}
	\label{sec:trade_study}
	\begin{figure}[h]
		\centering
		\includegraphics[width=\columnwidth]{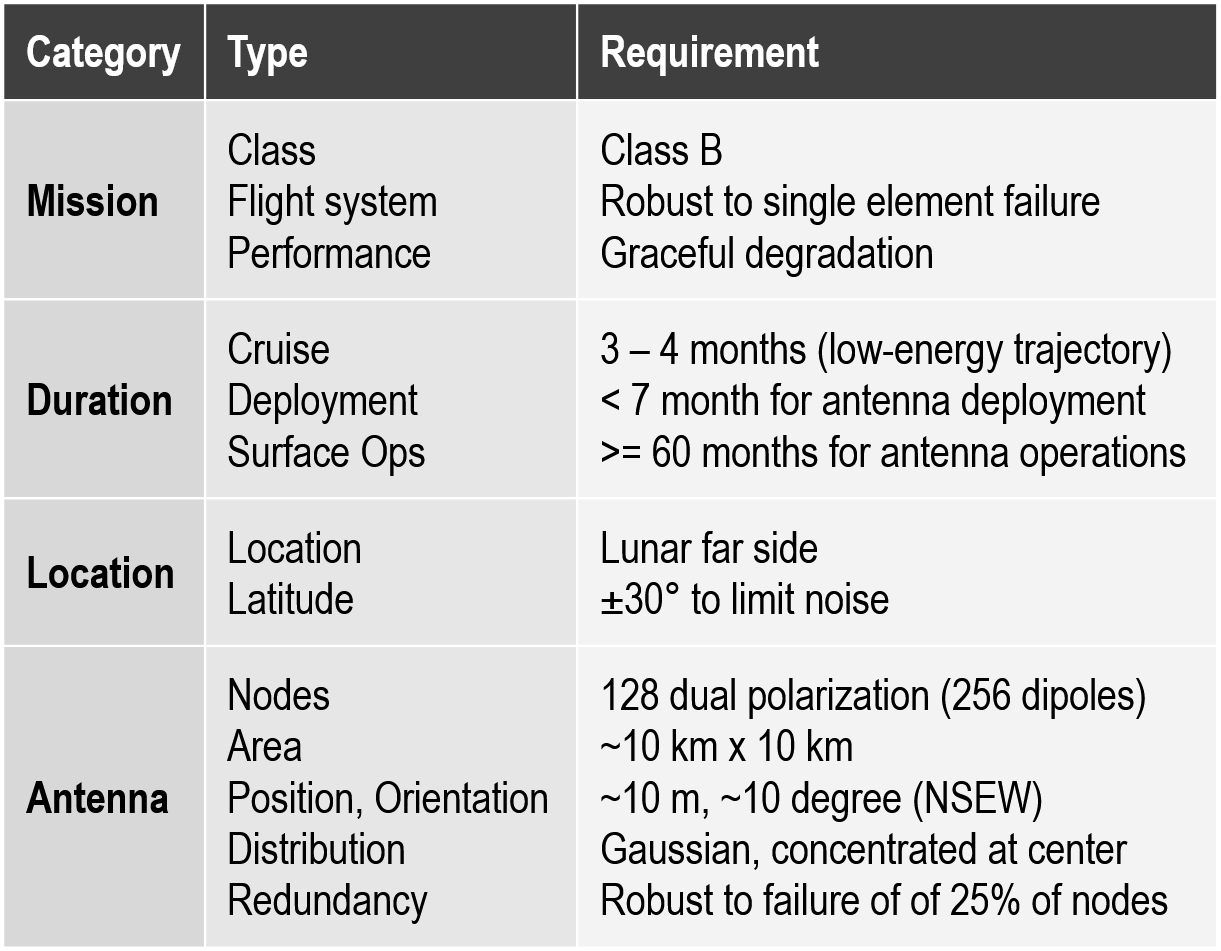}
		\caption{\emph{FARSIDE Requirements}}
		\label{fig:requirements}
	\end{figure}
	As described in Figure\,\ref{fig:requirements}, the trade study for the deployment of FARSIDE was driven by high-level requirements to meet science goals within the funding architecture of NASA's mission classes and risk posture. A number of factors were considered in the trade and are detailed as follows. 

	\subsection{Antenna Performance}
	\label{subsec:antenna}
	
\begin{figure}[h]
		\centering
		\includegraphics[width=\columnwidth]{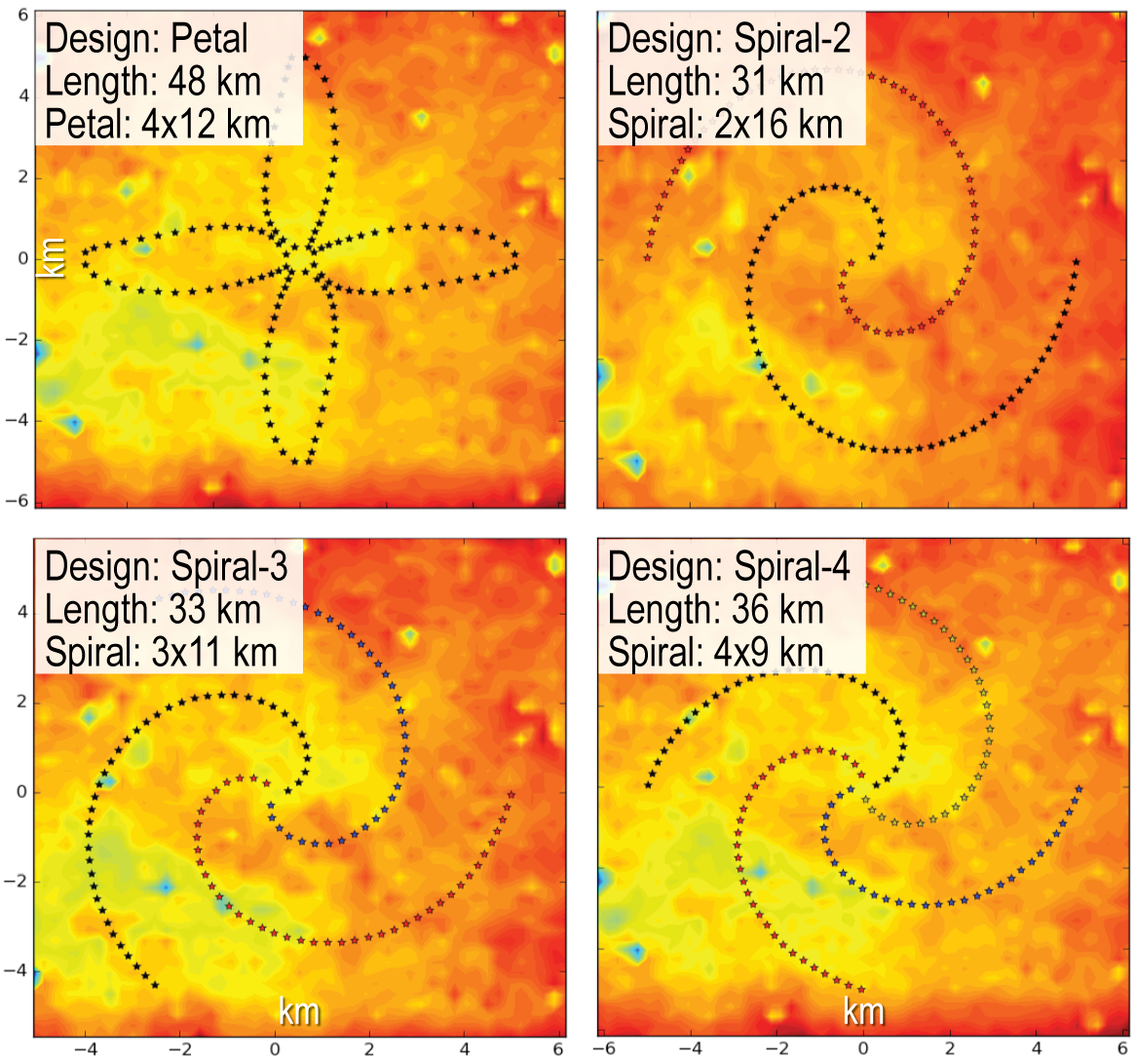}
		\includegraphics[width=\columnwidth]{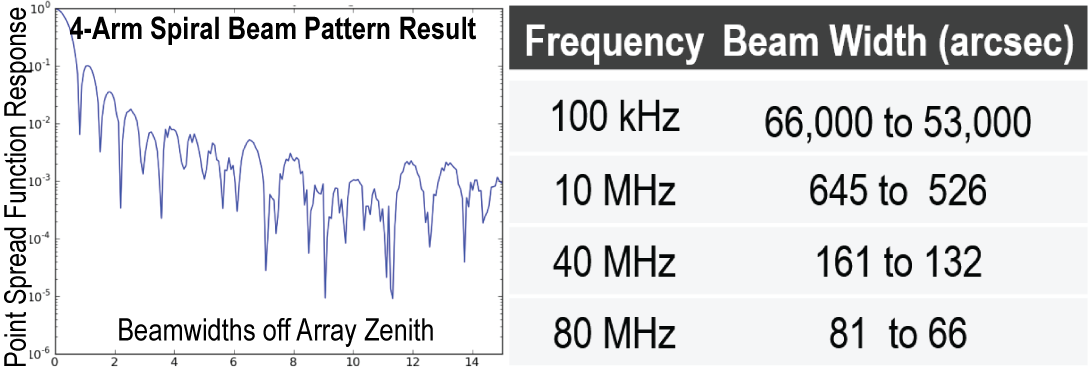}
		\caption{\emph{Topology Examples}: (\textit{Top}) Analysis was performed on four layout topologies overlaid on an elevation model of the lunar surface. (\textit{Bottom}) The point spread function (PSF) and beam width at various frequencies are shown for the favored 4-arm spiral approach. Note that the petal topology could be implemented using 1 to 8 rovers, but overall, performed slightly poorer than the spiral.}
		\label{fig:shape-compare}
	\end{figure}

	In reference to the ideal node distribution shown previously (Figure\,\ref{fig:ideal-node-distribution}), we first considered layout topologies  (i.e., patterns/shapes on the surface) that could best approximate the ideal, Gaussian-like distribution of nodes. However, to avoid the complexity of either \textit{i}) many rovers attached to tethers interfering with one another or \textit{ii}) a single rover with an extremely long tether, we instead favored an approach to determine the fewest number of rovers needed to deploy a tether along simple trajectories. Then we used beam function modeling to assess if that distribution performed well enough to achieve the science goals set forth in \cite{burns2021lunar}. For example, FARSIDE originally had proposed a single large rover carrying a continuous, 48\,km tether that would be deployed into a 4-petal, flower shape\,\cite{burns2019farside} (see ). In our updated analysis, we traded several topologies (including the petal) and concluded that, from both a science and deployment robustness standpoint, a spiral pattern was ideal. Favorable aspects of the spiral topology include the following.
	\begin{itemize}
		\item[\textcolor{teal}{$\textbf{+}$}] deployed with any number of spiral arms
		\item[\textcolor{teal}{$\textbf{+}$}] redundancy in case any rovers fail (for rover $>2$)
		\item[\textcolor{teal}{$\textbf{+}$}] one-way deployment reduces chance of tether fouling
		\item[\textcolor{teal}{$\textbf{+}$}] distributes mass and length of tether
	\end{itemize}
	Analysis was performed on four different topologies illustrated in Figure\,\ref{fig:shape-compare}, including the petal and 2-, 3-, and 4-arm spirals. The best performance and distribution is achieved by a 4-arm spiral (utilizing 4 rovers with 12\,km tethers) with negligible impact to the science capabilities. Additionally, this layout required the shortest rover trajectories, which implies easier accommodation of a smaller tether spool and less mass burden per rover.
	
\begin{figure}[h]
		\centering
		\includegraphics[width=0.8\columnwidth]{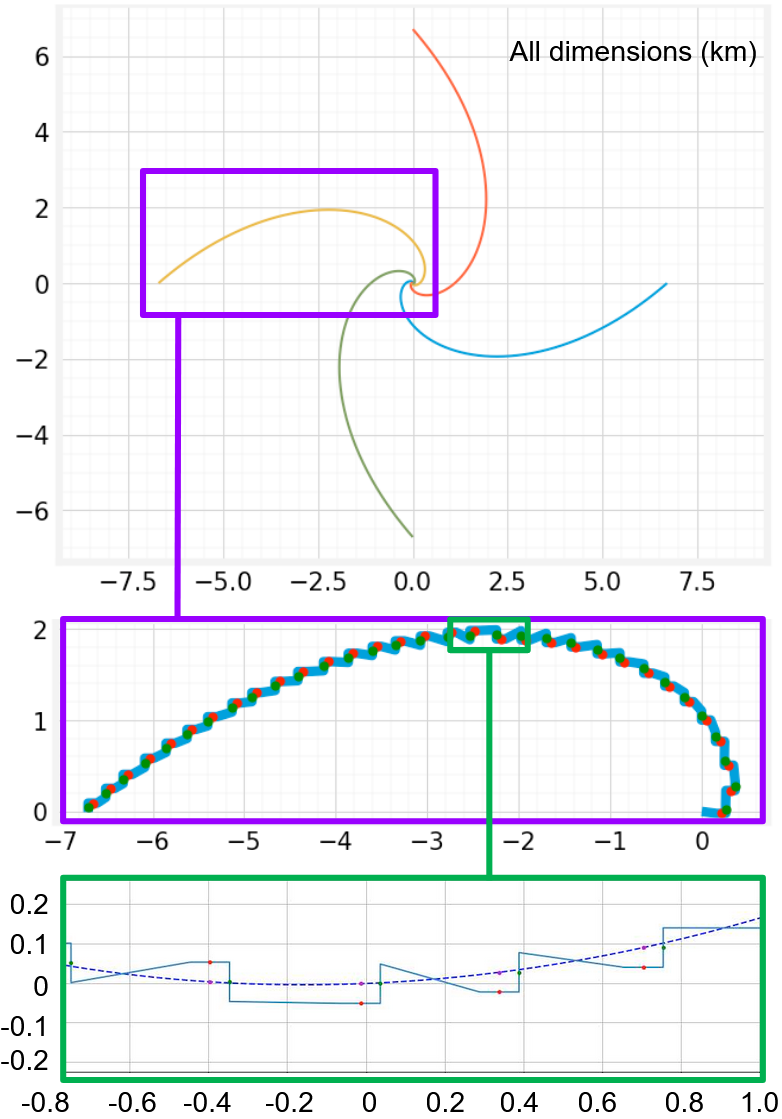}
		\caption{\emph{Local Layout Pattern}: The detailed layout of a tether along a 9\,km spiral arm  is shown. Given a requirement to layout antenna segments aligned to cardinal directions (NSEW), the total path and tether length are 12\,km.}
		\label{fig:design-zoom-in}
	\end{figure}

	Key to the operation of the network of dual-polarization nodes is their global orientation. No matter what layout topology is used, the deployment rover is required to execute a local `zig-zag' trajectory in order to layout antenna segments (embedded in a single tether) that are both orthogonal and aligned either North-South (NS) or East-West (EW). Thus, the total tether and trajectory distance are greater than the length of the spiral arm. Figure\,\ref{fig:design-zoom-in} shows what this looks like in detail. For example, a 9-km arm results in a tether that is 12\,km. In order to execute this complex path, the rover, which contains a tether spool, needs to payout tether with very low tension and make relatively tight orthogonal turns without disrupting the laid down tether.
	
	\begin{figure}[H]
		\centering
		\includegraphics[width=\columnwidth]{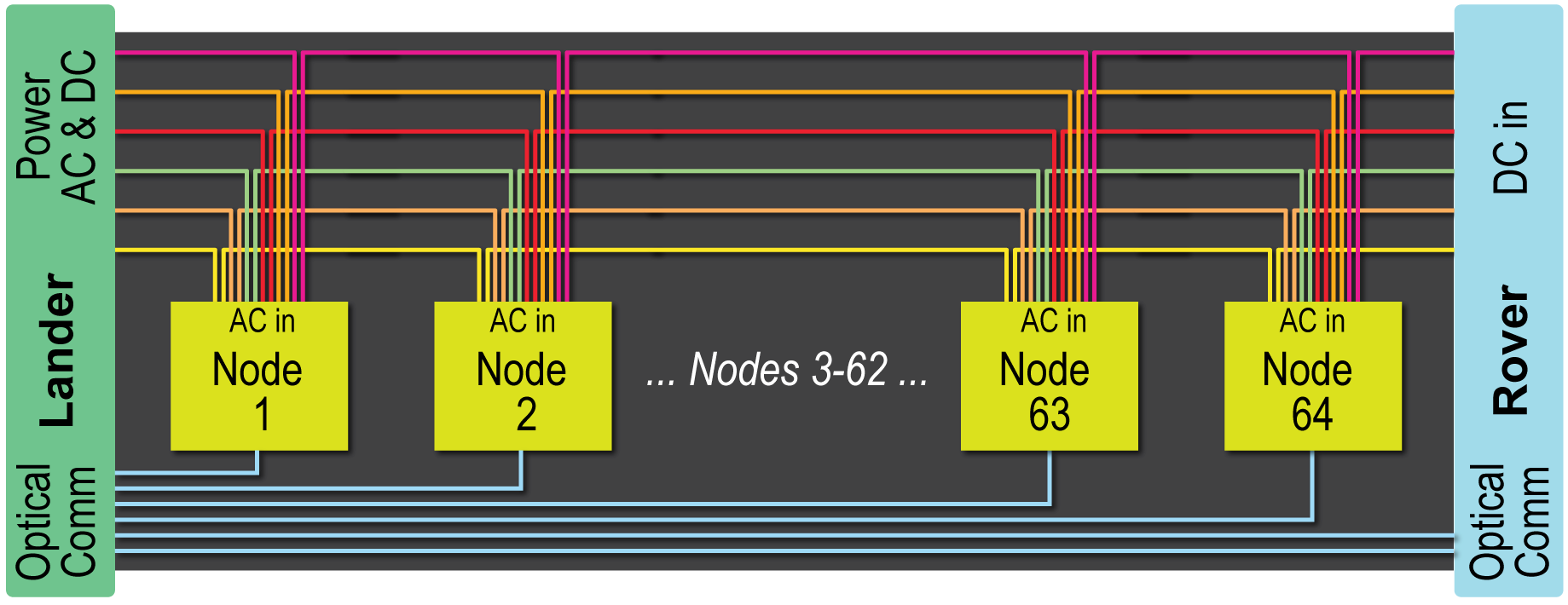}
		\caption{\emph{Tether Architecture}: An illustration shows the elements and connectivity of the tether. Each of 64 nodes (per tether and rover) is connected by a discrete single-mode, fiber-optic line and is powered using 6 redundant 28-AWG wires via 3-Phase AC. If a single node fails, similar to holiday tree lights, all other nodes will continue to be powered in parallel. The rover receives power from the same lines, but switched temporarily to DC to prevent unwanted operation of the AC power nodes during deployment. The rover can also communicate through the entire tether back to the lander for telemetry, control, and imaging purposes.}
		\label{fig:tether-arch}
	\end{figure}
	
	\vspace{-15px}
	
	\subsection{Antenna/Tether Design}
	\label{subsec:tether}
	
\begin{figure}[h]
		\centering
		\includegraphics[width=\columnwidth]{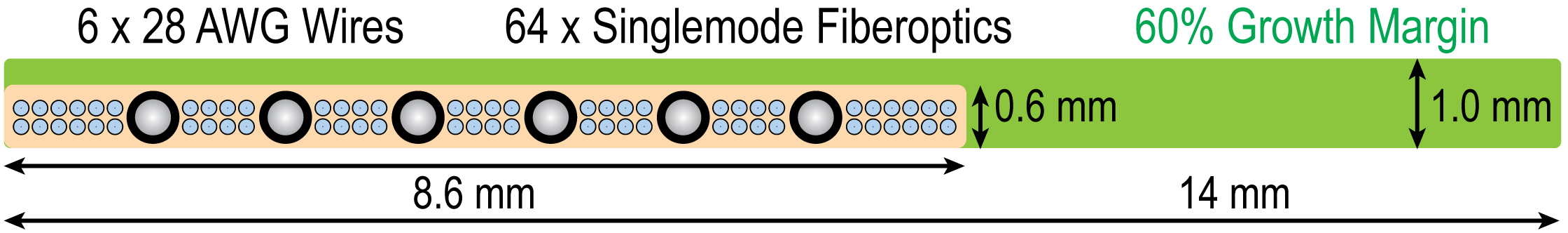}
		\caption{\emph{Tether Cross Section}: This true-scale drawing shows the actual diameters of single-mode fiberoptics and polyamide coated wires, which are stacked and arranged in a compact pattern to reduce the cable width and total height. Most importantly, we favored the thinnest tether in order to accommodate into a spool roll.}
		\label{fig:tether-cross-section}
	\end{figure}

	The design of the FARSIDE antenna/tether involves several novel capabilities listed as follows.
	\begin{itemize}
		\item[\textcolor{teal}{$\textbf{+}$}] 64 dipoles are embedded in 4 tethers (1 per rover)
		\item[\textcolor{teal}{$\textbf{+}$}] dipoles use the same wire as power wires
		\item[\textcolor{teal}{$\textbf{+}$}] embedded MnZn ferrite is used for dipole isolation
		\item[\textcolor{teal}{$\textbf{+}$}] 64 cold-cable electronic modules are appended
		\item[\textcolor{teal}{$\textbf{+}$}] fiberoptics connect every node back to the lander
		\item[\textcolor{teal}{$\textbf{+}$}] High voltage DC is used to power the rover
		\item[\textcolor{teal}{$\textbf{+}$}] AC is used to power the nodes after DC shutoff
	\end{itemize}
	An illustration of the architecture for such a tether is shown in Figure\,\ref{fig:tether-arch}.
	Given the constraints on tether design, we traded the overall cross-sectional design, as there are many ways to implement a tether or cable. Generally, they fall into the broader categories of circular or flat. Depending of the environment and application, there are pros and cons to each. However, for our case, the most important parameters for selection were the ease of integrating electronics nodes along the tether and its capabilities to be reliably unspooled (laid out) at low tension on the surface. Figure\,\ref{fig:tether-cross-section} captures the cross-sectional design trade in more detail.
	\begin{figure}[h]
		\centering
		\includegraphics[width=\columnwidth]{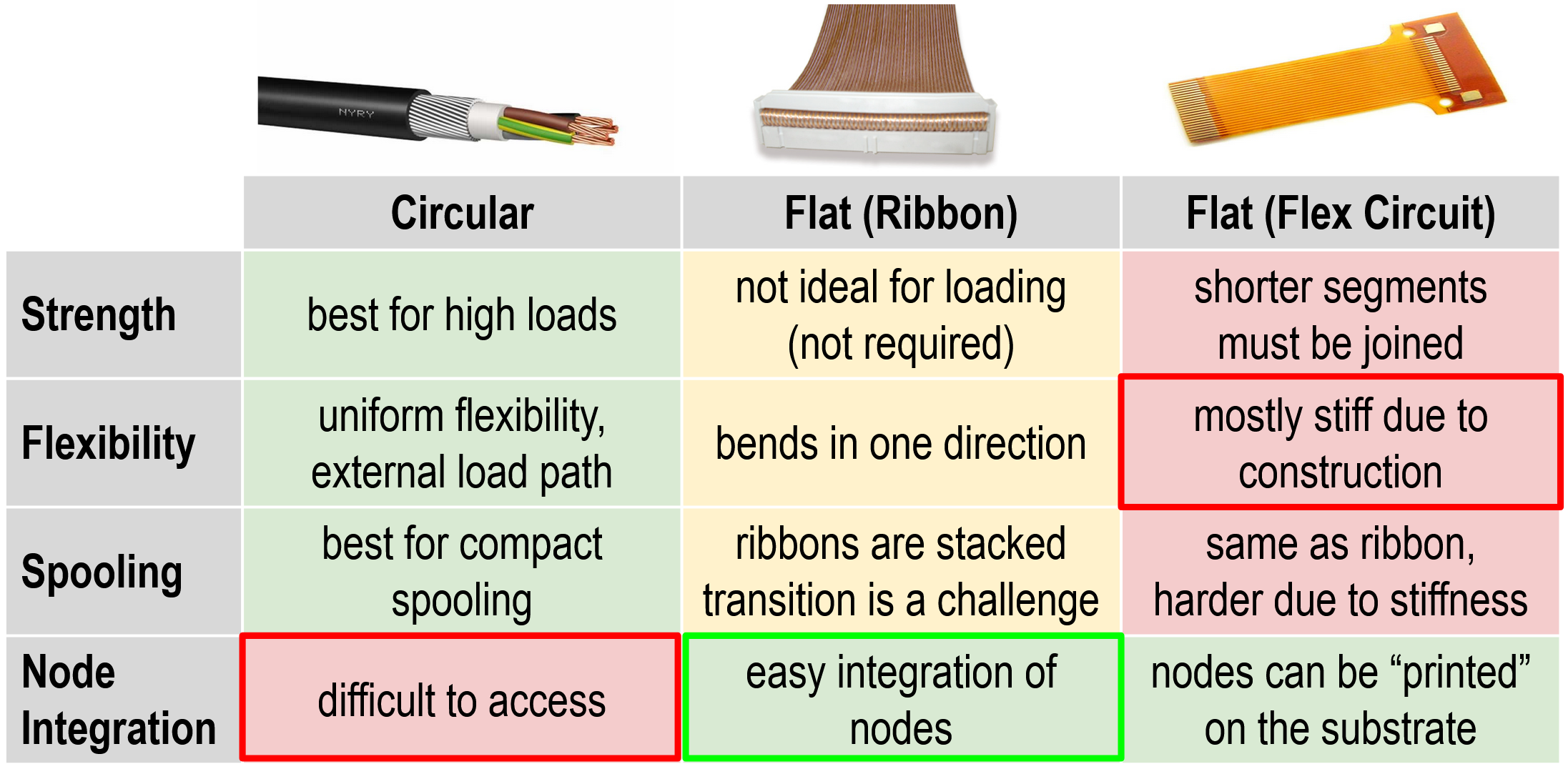}
		\caption{\emph{Cross-Sectional Design Trade}: Circular and flat cables are compared for the FARSIDE application. Overall, the key parameters for selection was the need to integrate 64 nodes and store them on a spool for easy deployment. Flat ribbon cables, similar to those used in the Apollo ALSEP experiment, were selected for their ability to be constructed at sufficiently long lengths on the order of kilometers.}
		\label{fig:tether-design-compare}
	\end{figure}
	
	Next, we determined a detailed cross-sectional allocation for 64 optical fibers and 6 28-AWG wires. The preferred configuration is shown in Figure\,\ref{fig:tether-cross-section}. Note, for rover and spool sizing purposes, we have included a 60\% margin to account for growth if the cable design cannot be achieved as shown. 
	\begin{figure}[h]
		\centering
		\includegraphics[width=0.9\columnwidth]{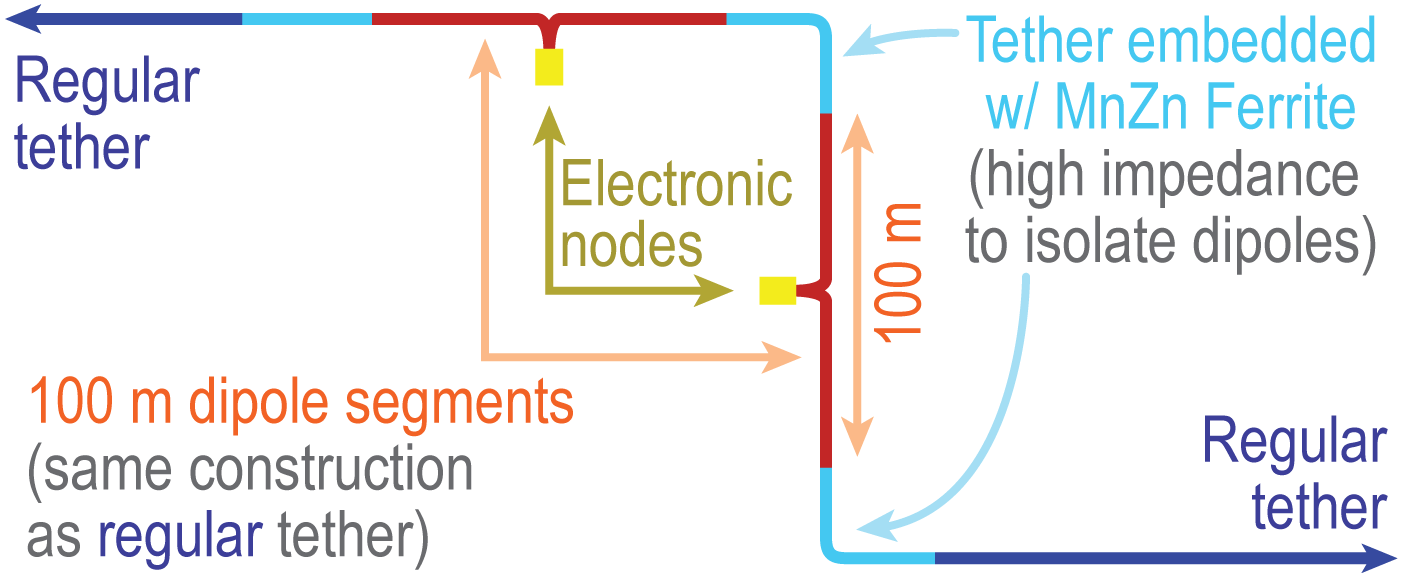}
		\caption{\emph{Length-Wise Tether Design}: The cable is shown with color-coded elements annotated. Note the high-impedance Manganese Zinc (MnZn) segments are used to isolate adjacent dipoles.}
		\label{fig:tether-segments}
	\end{figure}

	The length-wise tether design is shown in Figure\,\ref{fig:tether-segments}. The most important features are the dove tails joining the electronics nodes to the tape tether and the embedded Manganese Zinc (MnZn) wire shrouds that will provide high impedance between dipole elements in the wires for isolation purposes.

	\subsection{Rover \& Spool Design} %TODO power story
	\label{subsec:rover}
	
\begin{figure}[h]
		\centering
		\includegraphics[width=0.8\columnwidth]{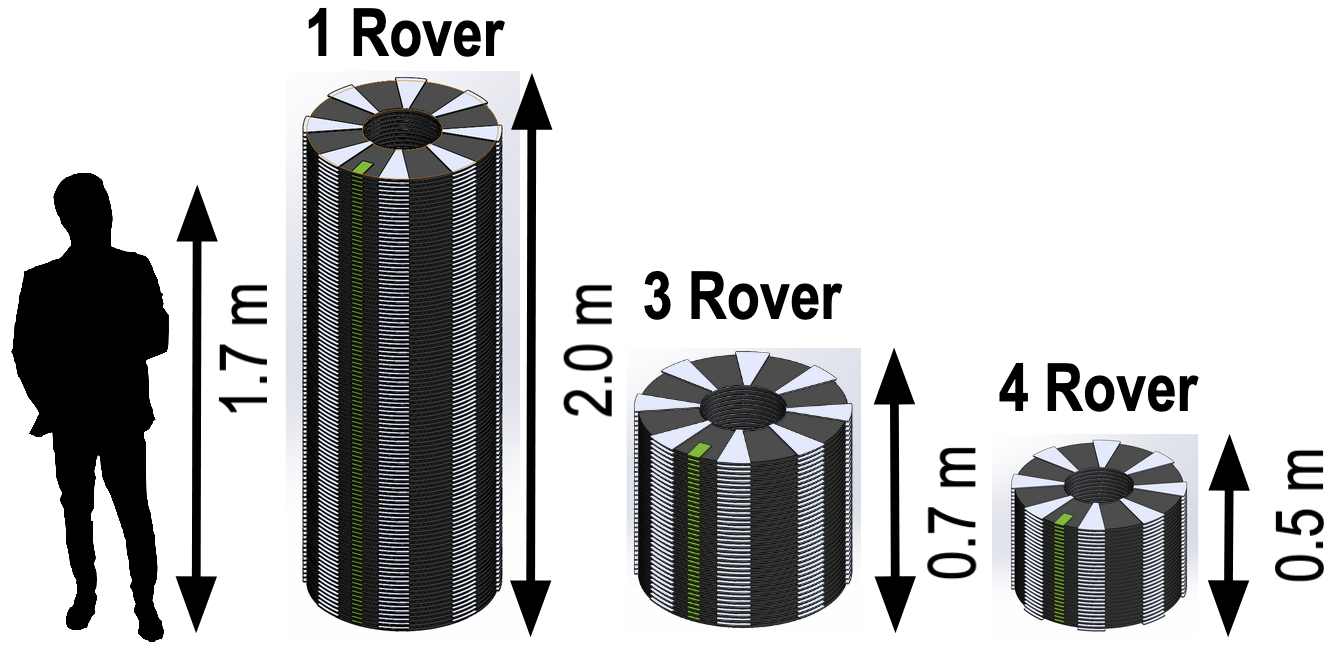}
		\includegraphics[width=0.8\columnwidth]{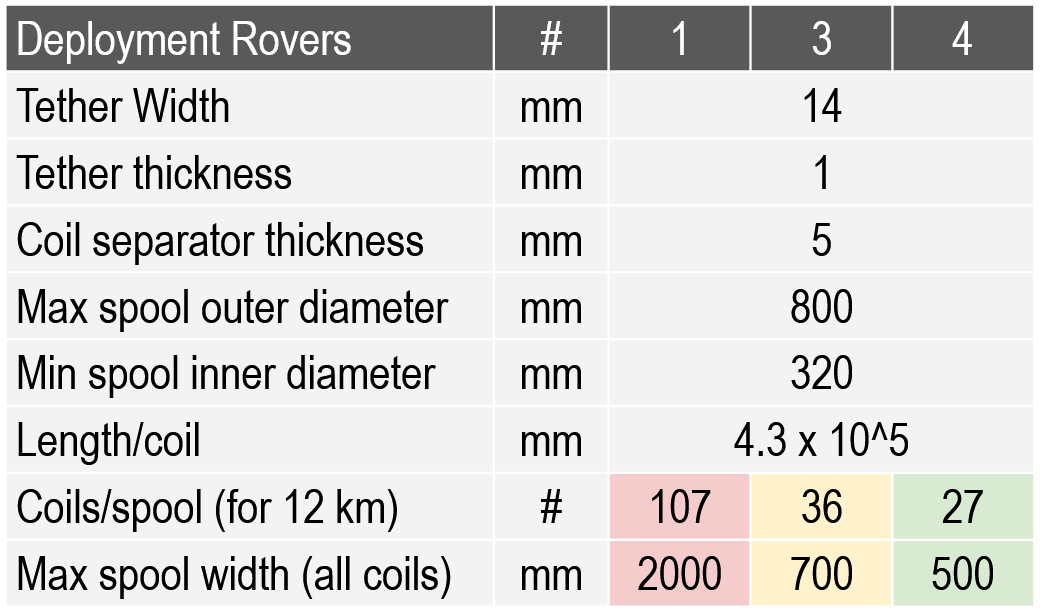}
		\caption{\emph{Spool Dimensions}: We compare the scale of tether spools for different number of deployment rovers assuming a minimum and maximum inner and outer spool diameter (allowing the width to grow as needed). In order to determine the number of spools to accommodate the total required length of tether, we used the following equation. Note, a coil is used in this context to mean a continuous roll of tether similar to the way gift ribbon is wound. The length for a single coil of the spool is $l=\pi t (d+h(t-1))$, where $t$ is the number of turns within the coil, $d$ is the outer coil diameter, and $h$ is the tether thickness. With the coil length calculated, we can then determine how many coils are stacked in the overall spool, and calculate its overall width.}
		\label{fig:spool-dims}
	\end{figure}

	The baseline design for the deployment mobility system is JPL's extreme-terrain Axel rover\,\cite{nesnas2012axel}. Axel has been developed over the last 20 years and is the most advanced tethered rover developed for space to date. Axel's simplistic design allows it to unspool and respool its tether (limiting drag and abrasion) from its central body and it achieves redundant mobility through just four degrees of freedom (wheels, body, ground reaction boom, and spool). Axel has been proposed, in various modified forms, for several mission concepts to explore the Moon, Mars, and beyond. For FARSIDE, the key driver for the Axel's size is the spool mass and geometry necessary to lay out a single spiral arm. As such, the typical approach is first to design a tether cross section, then a spool, and, finally, scale the rover.
	Thus, the deployment trade was directly coupled to the rover quantity given the need to accommodate continuous lengths of tether.
	Earlier, we showed the impact of the spiral design on length using different rover combinations. Figure\,\ref{fig:spool-dims} illustrates why the 4-rover system was preferred.

\begin{figure}[h]
		\centering
		\includegraphics[width=0.9\columnwidth]{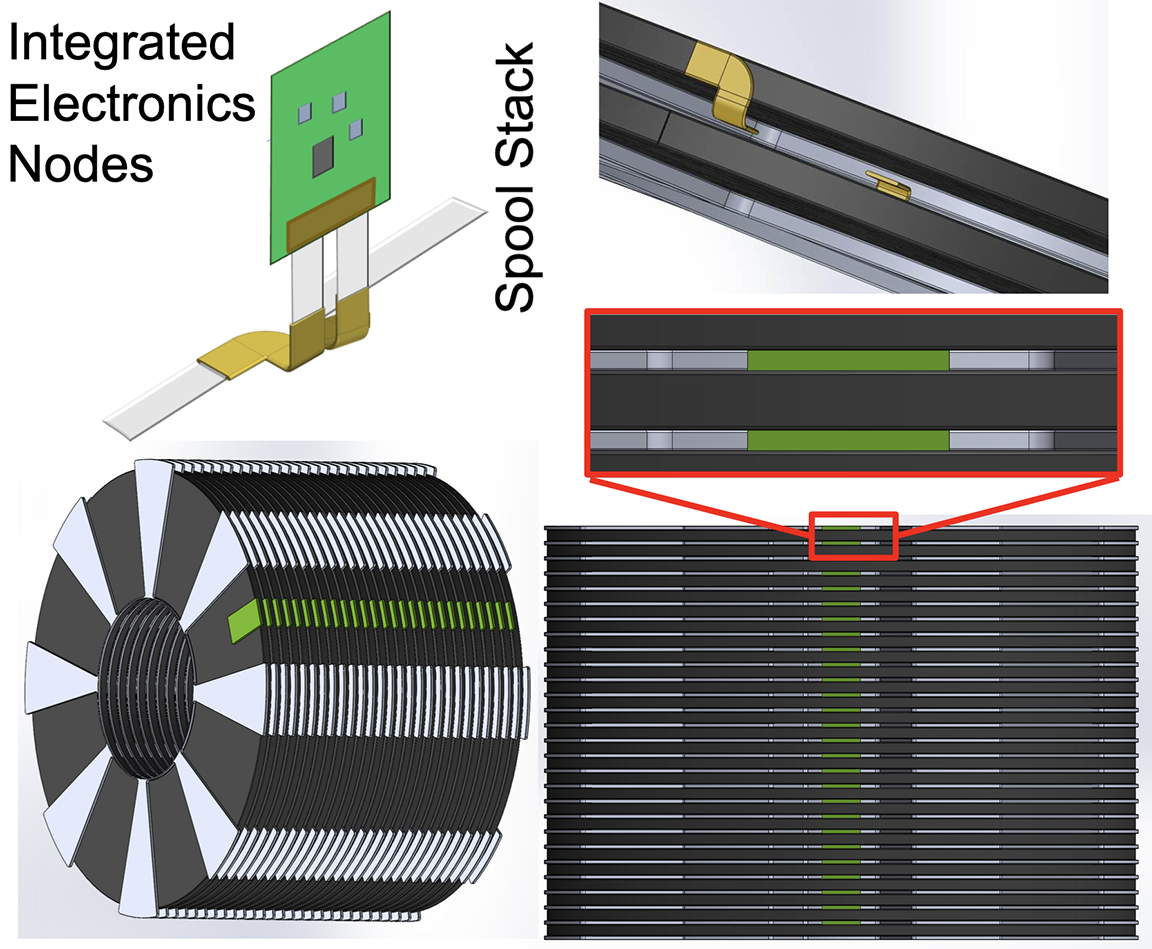}
		\caption{\emph{Spool Design}: A stack of coils with integrated electronics nodes are shown. Flexible dove tail connections are used to terminate the tether to each electronics node, which is safely stored at the outside diameter of the spool for roll off onto the surface.}
		\label{fig:spool-design}
	\end{figure}

	The selected spool design is best fit for a flat tether with protruding electronics nodes as shown in Figure\,\ref{fig:spool-design}. The operation of the spool allows a single coil to roll off at a time and provides a seamless transition to the next spool.

\begin{figure}[h]
		\centering
		\includegraphics[width=0.9\columnwidth]{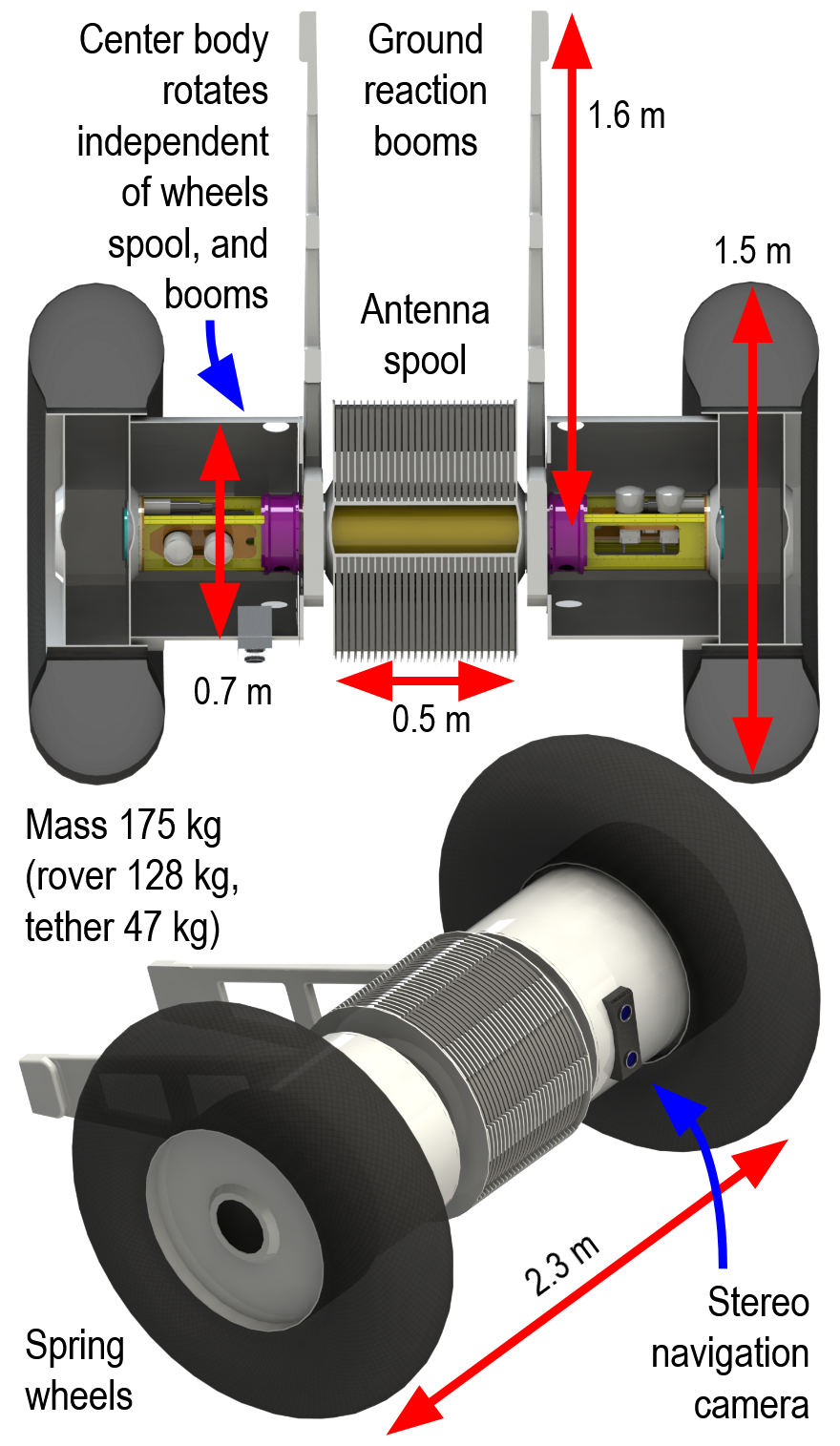}
		\caption{\emph{Axel Rover}: A rendering of the scaled FARSIDE Axel rover is shown. Each rover is 128\,kg and the total system including 12\,km of tether) is 175\,kg. Axel's average speed is $\sim$0.1\,m/s. Each spiral arm is expected to be deployed within a single lunar day (14 Earth days) with sufficient margin.}
		\label{fig:axel-rover}
	\end{figure}

	Finally, given the geometry of the spool, we scaled the Axel rover appropriately as shown in Figure\,\ref{fig:axel-rover}. A major modification to Axel for the FARSIDE mission is the use of two separately actuated booms on either side of its spool to provide a ground reaction force. The primary reason for this modification was to minimize interference of the tether with the booms and allow more flexibility with in-place turns for rover during the `zig-zag' tether deployment stage. By lifting the boom in the direction of the turn, the outer boom can still react the ground and tether will be deployed without significant dragging or chance of being run over by the inner boom. The orthogonal shape of the tether at the turn allows for some leeway and does not need to form a sharp angle. Also, given that the tether is flat, we do not expect that the tether at the point of turn will rest flat on the surface. In either case, the scale of the dipole, at 100\,mm, implies that some inconsistencies in its deployed shape and orientation are acceptable.
	
	\subsection{Lander Accommodation}
	\label{subsec:lander}
	
\begin{figure}[h]
		\centering
		\includegraphics[width=\columnwidth]{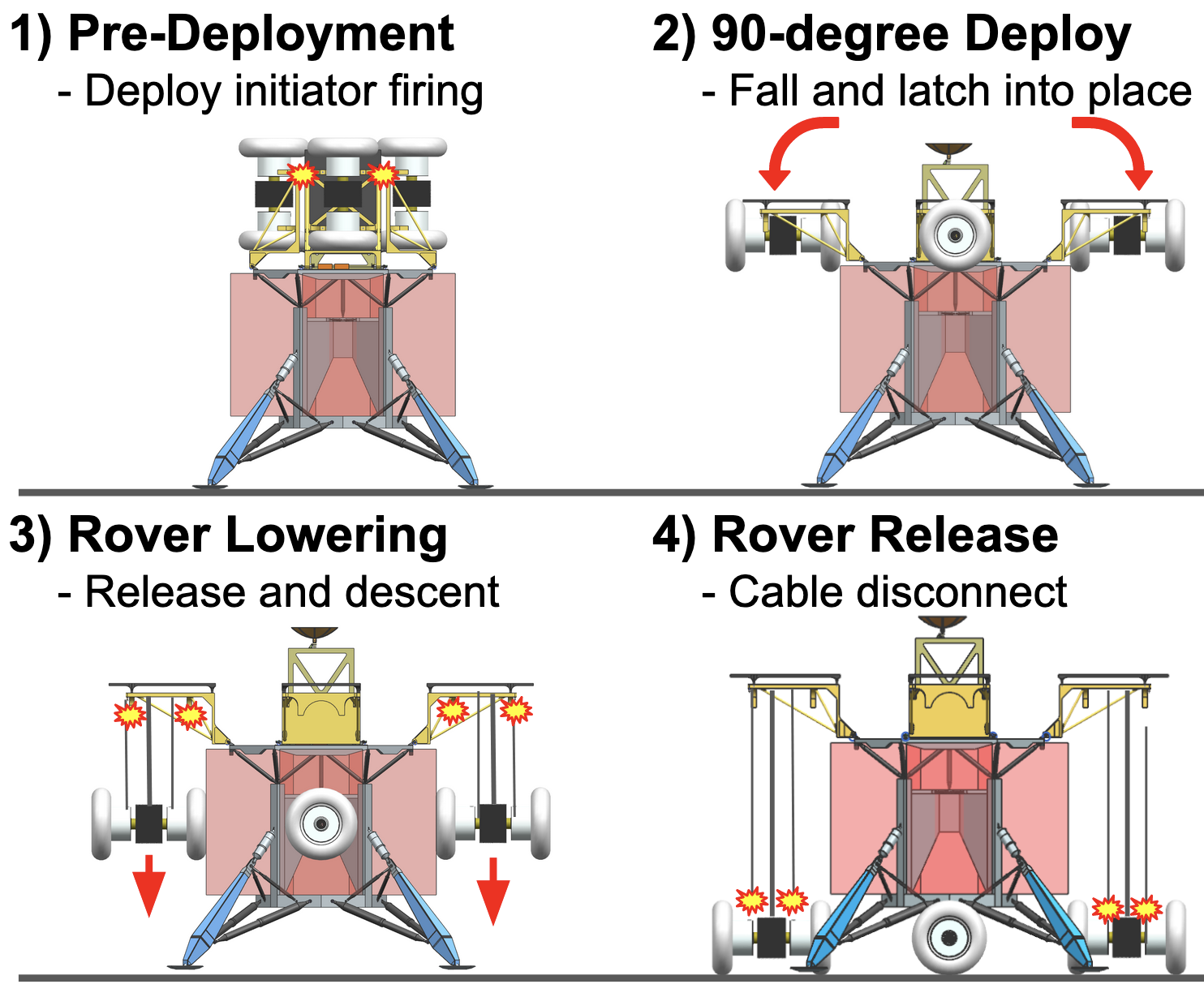}
		\caption{\emph{Lander Accommodation}: This sequence shows the stowed and egress states for each of four Axel rovers. The drawing is notional as the lander implementation and details are subject to change. The lander drawing and model were provided courtesy of Blue Origin.}
		\label{fig:egress}
	\end{figure}

	Using Blue Origin's Blue Moon Lander for a case study, we determined a compact, robust packing and deployment concept for 4 tethered rovers. The step-by-step approach is explained in Figure\,\ref{fig:egress}. Using a hinged \textit{Davit} system, similar to those used for rescue craft on large ships, Axel rovers are lowered to the surface on a pair of support only cables so that the tension on the science tether remains negligible. Once, on the surface, the support cables are released and each Axel can start roving. Further, the lander will house the integration computer and electronics to receive the data from each of the 256 nodes (via fiberoptics) and package it for efficient transmission to Earth via the Lunar Gateway.
	
	\subsection{Concept of Operation}
	\label{subsec:conops}
	
\begin{figure*}[hbtp]
		\centering
		\includegraphics[width=\textwidth]{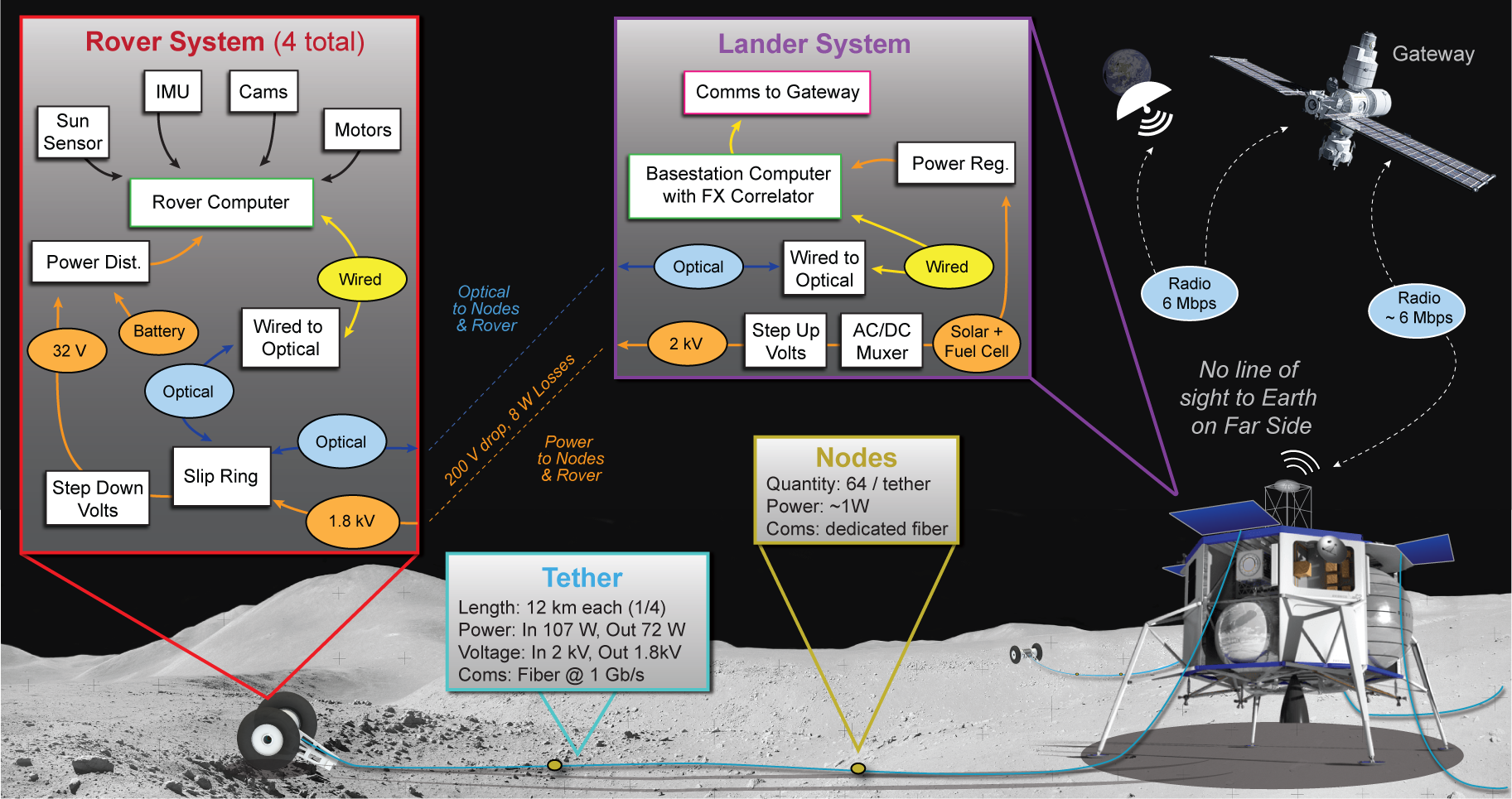}
		\caption{\emph{System Architecture and FARSIDE Concept of Operations}}
		\label{fig:system-arch}
	\end{figure*}

	Finally, Figure\,\ref{fig:system-arch} on Page 7, explains the overall system architecture and concept of operations at a high level.  Note that NASA's planned Lunar Gateway is the baseline plan for communication to the FARSIDE system given non-line-of-sight position with respect to Earth. An alternative would be to launch a support satellite along with FARSIDE. 
	
	%%%%%%%%%%%%%%%%%%%%%%%%%%%%%%%%%%%%%%
	
	\section{Conclusion \& Next Steps}
	\label{sec:Conclusion}
	
	We have summarized a trade study for a mission concept to deploy a 10-km wide interferometer radio telescope on the Moon with Just 4 tethered rovers. The feasibility of the study is based on analysis and simulation of the antenna performance as well as mobility and tethered operations of a previously fielded Axel rover system (we did not do any demonstrations of the FARSIDE system). Future work will require vetting and testing all the approaches discussed as well as development of lower TRL technology. FARSIDE is a mission concept and is thus in early development stages and not yet selected for flight.
	
	%%%%%%%%%%%%%%%%%%%%%%%%%%%%%%%%%%%%%%
	
	\acknowledgments
	
	This research was carried out at the Jet Propulsion Laboratory, California Institute of Technology, under a contract with the National Aeronautics and Space Administration. The information presented about the mission concept is pre-decisional and is provided for planning and discussion purposes only.
	
	%%%%%%%%%%%%%%%%%%%%%%%%%%%%%%%%%%%%%%

%	\balance
	\interlinepenalty=10000
	\bibliographystyle{IEEEtran}
	\bibliography{refs}

\begin{thebibliography}{10}
\providecommand{\url}[1]{#1}
\csname url@samestyle\endcsname
\providecommand{\newblock}{\relax}
\providecommand{\bibinfo}[2]{#2}
\providecommand{\BIBentrySTDinterwordspacing}{\spaceskip=0pt\relax}
\providecommand{\BIBentryALTinterwordstretchfactor}{4}
\providecommand{\BIBentryALTinterwordspacing}{\spaceskip=\fontdimen2\font plus
\BIBentryALTinterwordstretchfactor\fontdimen3\font minus
  \fontdimen4\font\relax}
\providecommand{\BIBforeignlanguage}[2]{{%
\expandafter\ifx\csname l@#1\endcsname\relax
\typeout{** WARNING: IEEEtran.bst: No hyphenation pattern has been}%
\typeout{** loaded for the language `#1'. Using the pattern for}%
\typeout{** the default language instead.}%
\else
\language=\csname l@#1\endcsname
\fi
#2}}
\providecommand{\BIBdecl}{\relax}
\BIBdecl

\bibitem{rajan2016space}
R.~T. Rajan, A.-J. Boonstra, M.~Bentum, M.~Klein-Wolt, F.~Belien, M.~Arts,
  N.~Saks, and A.-J. van~der Veen, ``Space-based aperture array for ultra-long
  wavelength radio astronomy,'' \emph{Experimental Astronomy}, vol.~41, no.
  1-2, pp. 271--306, 2016.

\bibitem{miralda2003dark}
J.~Miralda-Escud{\'e}, ``The dark age of the universe,'' \emph{Science}, vol.
  300, no. 5627, pp. 1904--1909, 2003.

\bibitem{burns1989lunar}
J.~O. Burns, N.~Duric, S.~Johnson, and G.~J. Taylor, ``A lunar far-side very
  low frequency array: suggestions for future work.'' \emph{Lunar Far-Side Very
  Low Frequency Array}, pp. 77--78, 1989.

\bibitem{burns1990lunar}
J.~O. Burns, ``The lunar observer radio astronomy experiment (lorae),'' in
  \emph{Low frequency astrophysics from space}.\hskip 1em plus 0.5em minus
  0.4em\relax Springer, 1990, pp. 19--28.

\bibitem{lazio2009dark}
T.~J.~W. Lazio, J.~Burns, D.~Jones, J.~Kasper, S.~Neff, R.~MacDowall, K.~Weiler
  \emph{et~al.}, ``The dark ages lunar interferometer (dali) and the radio
  observatory for lunar sortie science (rolss),'' in \emph{American
  Astronomical Society Meeting Abstracts\# 213}, vol. 213, 2009, pp. 451--02.

\bibitem{lazio2011radio}
T.~J.~W. Lazio, R.~MacDowall, J.~O. Burns, D.~Jones, K.~Weiler, L.~Demaio,
  A.~Cohen, N.~P. Dalal, E.~Polisensky, K.~Stewart \emph{et~al.}, ``The radio
  observatory on the lunar surface for solar studies,'' \emph{Advances in Space
  Research}, vol.~48, no.~12, pp. 1942--1957, 2011.

\bibitem{burns2011dark}
J.~O. Burns, J.~Lazio, J.~Bowman, R.~Bradley, C.~Carilli, S.~Furlanetto,
  G.~Harker, A.~Loeb, and J.~Pritchard, ``The dark ages radio explorer
  (dare),'' in \emph{American Astronomical Society Meeting Abstracts\# 217},
  vol. 217, 2011, pp. 107--09.

\bibitem{smith2020artemis}
M.~Smith, D.~Craig, N.~Herrmann, E.~Mahoney, J.~Krezel, N.~McIntyre, and
  K.~Goodliff, ``The artemis program: An overview of nasa's activities to
  return humans to the moon,'' in \emph{2020 IEEE Aerospace Conference}.\hskip
  1em plus 0.5em minus 0.4em\relax IEEE, 2020, pp. 1--10.

\bibitem{nesnas2012axel}
I.~A. Nesnas, J.~B. Matthews, P.~Abad-Manterola, J.~W. Burdick, J.~A. Edlund,
  J.~C. Morrison, R.~D. Peters, M.~M. Tanner, R.~N. Miyake, B.~S. Solish
  \emph{et~al.}, ``Axel and duaxel rovers for the sustainable exploration of
  extreme terrains,'' \emph{Journal of Field Robotics}, vol.~29, no.~4, pp.
  663--685, 2012.

\bibitem{mcgarey2020design}
P.~McGarey, T.~Nguyen, T.~Pailevanian, and I.~Nensas, ``Design and test of an
  electromechanical rover tether for the exploration of vertical lunar pits,''
  in \emph{2020 IEEE Aerospace Conference}.\hskip 1em plus 0.5em minus
  0.4em\relax IEEE, 2020, pp. 1--10.

\bibitem{bandyopadhyay2018conceptual}
S.~Bandyopadhyay, J.~Lazio, A.~Stoica, P.~Goldsmith, B.~Blair, M.~Quadrelli,
  J.-P. de~la Croix, and A.~Rahmani, ``Conceptual ideas for radio telescope on
  the far side of the moon,'' in \emph{2018 IEEE Aerospace Conference}.\hskip
  1em plus 0.5em minus 0.4em\relax IEEE, 2018, pp. 1--10.

\bibitem{burns2019farside}
J.~Burns, G.~Hallinan, J.~Lux, A.~Romero-Wolf, T.-C. Chang, J.~Kocz, J.~Bowman,
  R.~MacDowall, J.~Kasper, R.~Bradley \emph{et~al.}, ``Farside: a low radio
  frequency interferometric array on the lunar farside,'' \emph{arXiv preprint
  arXiv:1907.05407}, 2019.

\bibitem{burns2021lunar}
J.~Burns, G.~Hallinan, T.-C. Chang, M.~Anderson, J.~Bowman, R.~Bradley,
  S.~Furlanetto, A.~Hegedus, J.~Kasper, J.~Kocz \emph{et~al.}, ``A lunar
  farside low radio frequency array for dark ages 21-cm cosmology,''
  \emph{arXiv preprint arXiv:2103.08623}, 2021.

\bibitem{bates1979alsep}
J.~R. Bates, W.~Lauderdale, and H.~Kernaghan, \emph{ALSEP termination
  report}.\hskip 1em plus 0.5em minus 0.4em\relax National Aeronautics and
  Space Administration, Scientific and Technical~…, 1979, vol. 1036.

\bibitem{simmons197315}
G.~Simmons, D.~Strangway, P.~Annan, and R.~Baker, ``15. surface electrical
  properties experiment,'' \emph{Apollo 17: Preliminary Science Report}, vol.
  330, 1973.

\bibitem{grimm2018new}
R.~E. Grimm, ``New analysis of the apollo 17 surface electrical properties
  experiment,'' \emph{Icarus}, vol. 314, pp. 389--399, 2018.

\end{thebibliography}
	
	%%%%%%%%%%%%%%%%%%%%%%%%%%%%%%%%%%%%%%
	\thebiography
	%%%%%%%%%%%%%%%%%%%%%%%%%%%%%%%%%%%%%%
	
	\begin{biographywithpic}
		{Patrick McGarey}{bio_mcgarey.jpg} %% photo size 1in X 1.25in
		Ph.D. is a Robotics Technologist at JPL in the Robotic Mobility Group. Patrick received his Ph.D. in Aerospace Engineering from the University of Toronto, where he was a visiting Fulbright Scholar. His research is focused on the development of tethered systems and autonomy functions for the exploration of extreme environments. 
	\end{biographywithpic}

	\begin{biographywithpic}
		{Issa A. Nesnas}{bio_nesnas.jpg} %% photo size 1in X 1.25in
		Ph.D. is a Principal  Robotics Technologist and the former supervisor of the Robotic Mobility group at the Jet Propulsion Laboratory with over two decades of research in space robotics and automation. He is the principal investigator for the Axel robot and leads research in autonomy and mobility with a focus on extreme terrain.
	\end{biographywithpic}

	\begin{biographywithpic}
		{Adarsh Rajguru}{bio_rajguru.jpg} %% photo size 1in X 1.25in
		is a Mechanical Engineer at NASA Jet Propulsion Laboratory in Pasadena, California. Previously, Adarsh was an Engineering Intern at Masten Space Systems. Currently, he works on a variety of projects within the Mechanical Engineering section at JPL. Adarsh helped design and implement the rover spool and egress system.
	\end{biographywithpic}

	\begin{biographywithpic}
		{Matthew Bezkrovny}{bio_bezkrovny.jpg} %% photo size 1in X 1.25in 
		is a Mechanical Engineer at NASA Jet Propulsion Laboratory in Pasadena, California. He is an expert in spacecraft structures, modeling, and design. For this project, he was responsible for lander-side accommodation (onto a Blue Origin \textit{Blue Moon Lander}) and egress methods. He determined a method for the swing \textit{Davit} system.
	\end{biographywithpic}

	\begin{biographywithpic}
		{Vahraz Jamnejad}{bio_jamnejad.jpg} %% photo size 1in X 1.25in
		Ph.D. is a principle scientist at NASA Jet Propulsion Laboratory in Pasadena, California, where is is an experts in electrical engineering, electromagnetic, optics, and mathematics. For this project, Vahraz interpreted beam size models for an interferometer network and is the point-of-contact for performance modeling for FARSIDE.
	\end{biographywithpic}

	\begin{biographywithpic}
		{Jim Lux}{bio_lux.jpg} %% photo size 1in X 1.25in
		is project manager at NASA Jet Propulsion Laboratory and is currently working on the Sun Radio Interferometer Space Experiment (SunRISE), which is an array of six CubeSats operating as one very large radio telescope. For this project, Jim provided oversight into the tether design approach and implementation of deployment systems. 
	\end{biographywithpic}

	\begin{biographywithpic}
		{Eric Sunada}{bio_sunada.jpg} %% photo size 1in X 1.25in
		is the lead thermal technologist for JPL’s propulsion, thermal and materials engineering section. He has 30 years of spacecraft thermal engineering experience supporting missions. Sunada’s interests are in the development of enabling technologies for JPL’s spacecraft to extreme environments. Eric provided FARSIDE Thermal models.  
	\end{biographywithpic}

	\begin{biographywithpic}
		{Lawrence Teitelbaum}{bio_teitelbaum.jpg} %% photo size 1in X 1.25in
		Ph.D. is the Deep Space Network (DSN) Science Manager at NASA Jet Propulsion Laboratory. Lawrence's background is in RF microwave engineering and he served as the study lead for the JPL deployment study for FARSIDE, interfacing between JPL, University of Colorado Boulder, and Blue Origin. 
	\end{biographywithpic}

	\begin{biographywithpic}
		{Alexander Miller}{bio_miller.jpg} %% photo size 1in X 1.25in
		Ph.D. works closely with lander subsystem experts to define and manage payload interfaces, works with customers to derive payload-specific requirements, leads CAD packaging exercises, and facilitates design reference mission studies. Alex received his Ph.D. in Exploration Systems Design from the Arizona State University School of Earth and Space Exploration.
	\end{biographywithpic}

	\begin{biographywithpic}
		{Steve W. Squyres}{bio_squyres.jpg} %% photo size 1in X 1.25in
		Ph.D. is an American astronomer and planetary scientist. He was the principal investigator of the Mars Exploration Rover Mission (MER). He is the recipient of the 2004 Carl Sagan Memorial Award and the 2009 Carl Sagan Medal for Excellence in Communication in Planetary Science. He is chief scientist for Blue Origin. 
	\end{biographywithpic}

	\begin{biographywithpic}
		{Gregg Hallinan}{bio_hallinan.jpg} %% photo size 1in X 1.25in
		Ph.D. is a Professor in the Astronomy Department at Caltech and Director of the Owens Valley Radio Observatory and serves as Deputy PI for the FARSIDE mission concept. His interests are in magnetic activity in stellar and planetary systems and the search for radio transients. He is responsible for ensuring that the deployment approach enables the critical science goals. 
	\end{biographywithpic}

	\begin{biographywithpic}
		{Alex Hegedus}{bio_hegedus.jpg} %% photo size 1in X 1.25in 
		Ph.D. is a leading expert in simulating and interpreting measurements from space-based low frequency radio arrays.  He is a part of several concept array teams that are at various stages of development including SunRISE, RELIC, PRIME, Synchrotron, and FARSIDE. Alex provided simulations of the array to determine key parameters.
	\end{biographywithpic}

	\begin{biographywithpic}
		{Jack Burns}{bio_burns.jpg} %% photo size 1in X 1.25in 
		Ph.D. is a Professor in the Department of Astrophysical and Planetary Sciences and a Professor in the Department of Physics, both at the University of Colorado (CU) Boulder, and is Vice President Emeritus for Academic Affairs and Research for the CU System. Jack is the PI of the FARSIDE mission concept and leads overall development.
	\end{biographywithpic}

\end{document}